# Surface sensitivity of Rayleigh anomalies in metallic nanogratings

**Silvio Savoia,**[1,5,*] **Armando Ricciardi,**[2,5] **Alessio Crescitelli,**[2] **Carmine Granata,**[3] **Emanuela Esposito,**[4] **Vincenzo Galdi,**[1] **and Andrea Cusano**[2]

[1] *Waves Group, Department of Engineering, University of Sannio, I-82100, Benevento, Italy*
[2] *Optoelectronics Division, Department of Engineering, University of Sannio, I-82100, Benevento, Italy*
[3] *Institute of Cybernetics, "E. Caianiello," National Council of Research, I-80078 Pozzuoli (NA), Italy*
[4] *Institute for Microelectronics and Microsystems, National Council of Research, Department of Naples, I-80131 Napoli, Italy*
[5] *these authors contributed equally to this work*
[*] *sisavoia@unisannio.it*

**Abstract:** Sensing schemes based on Rayleigh anomalies (RAs) in metal nanogratings exhibit an impressive *bulk* refractive-index sensitivity determined solely by the grating period. However, the *surface* sensitivity (which is a key figure of merit for label-free chemical and biological sensing) needs to be carefully investigated to assess the actual applicability of this technological platform. In this paper, we explore the sensitivity of RAs in metal nanogratings when *local* refractive-index changes are considered. Our studies reveal that the surface sensitivity deteriorates up to two orders of magnitude by comparison with the corresponding bulk value; interestingly, this residual sensitivity is not attributable to the wavelength shift of the RAs, which are completely insensitive to local refractive-index changes, but rather to a strictly connected *plasmonic* effect. Our analysis for increasing overlay thickness reveals an ultimate surface sensitivity that approaches the RA bulk value, which turns out to be the upper-limit of grating-assisted surface-plasmon-polariton sensitivities.

## 1. Introduction

Spectral anomalies in the response of diffraction gratings constitute a subject of longstanding interest in optics, which originated from the observations by Wood in 1902 [1] and their physical interpretations provided by Rayleigh [2,3] and Fano [4]. The reader is referred to a recent book chapter by Maystre [5] (and references therein) for a nice historical perspective and a comprehensive review of the quantitative phenomenological theory. Basically, two types of anomalies can be identified: *sharp* anomalies, due to the passing-off of a spectral diffraction order, and *diffuse* anomalies arising from the excitation of surface waves. The former, typically referred to as Rayleigh anomalies (RAs), occur at wavelengths $\lambda_R^{(m)}$ given by the well-known grating formula [5]

$$\lambda_R^{(m)} = \frac{\Lambda n_a (-\sin\theta_i \pm 1)}{m}, \ m = \pm 1, \pm 2, ..., \qquad (1)$$

where a one-dimensional (1-D) grating of period $\Lambda$ immersed in a medium with refractive index (RI) $n_a$ is assumed, and $\theta_i$ denotes the angle of incidence (measured anticlockwise from the normal to the grating).

Examples of diffuse anomalies are *surface plasmon polaritons* (SPPs) occurring when the collective charge oscillations on the surface of a metallic grating are excited, or *guided-mode resonances*, which may occur, for instance, in dielectric-coated metallic gratings [6]. Both types of anomalies (sharp and diffuse) strongly affect the transmission/reflection of light through/by sub-wavelength metal gratings. They typically manifest themselves as rapid intensity variations in the spectra, and, depending on the type of periodic structure, may occur either separately or jointly (almost superimposed) [7].

It is worth pointing out that RAs and SPPs have very different near-field distributions. More specifically, for RAs, the field enhancement extends far from the grating surface, whereas, for SPPs, the field distribution is exponentially bound to the grating surface. Although both RA and SPPs are inherently sensitive to changes in the surrounding RI, this distinction bears important implications in connection with possible applications of these gratings to sensing, as it will be discussed hereafter.

An approach for using RA-based nanogratings (specifically, periodic hole arrays in thin gold films) for RI sensing was proposed in [8], where it was essentially demonstrated that the coupling of an SPP with an RA may lead to a large increase in the transmission of a narrow linewidth peak over a small RI range.

Based on the observation that the Rayleigh wavelengths in Eq. (1) are sensitive to changes of the surrounding ambient RI, a novel sensing scheme was proposed in a recent work [9]. Essentially, this approach is based on the monitoring of the wavelength shift of a reflection peak associated with an RA of a concentric gold ring nanograting laid on the end facet of an optical fiber. Among the potential advantages of this type of sensors, the authors highlight the polarization insensitivity (due to the rotational symmetry of the grating), the linear relationship between the RA-peak wavelength and the ambient RI [cf. Eq. (1)], with a *bulk* sensitivity determined solely by the grating period, namely 900 nm per RI unit (RIU). Given these impressive properties, one may be led to consider RA-based grating nanosensors as a valid alternative to standard SPP-based sensors [10,11] in label-free chemical and biological sensing applications. However, in these typical scenarios of practical interest, the RI changes are mainly restricted to surface modifications occurring at the sensor interface (e.g., due to the binding of a biological layer, physical or chemical adsorption of analyte molecules). Consistently with the inherently *diffractive* character of the RA phenomena (with *extended* field distributions) [12], the figures of merit predicted in [9] are expected to deteriorate when local (rather than bulk) RI changes are considered, as typical in all sensing schemes when the sensing volume is not entirely filled up. However, a *quantitative* assessment of such deterioration cannot be attained via simple considerations, since local surface modifications significantly enrich and complicate the underlying phenomenology. Therefore, an in-depth investigation of the optical response of the structure is required, which motivates our present study.

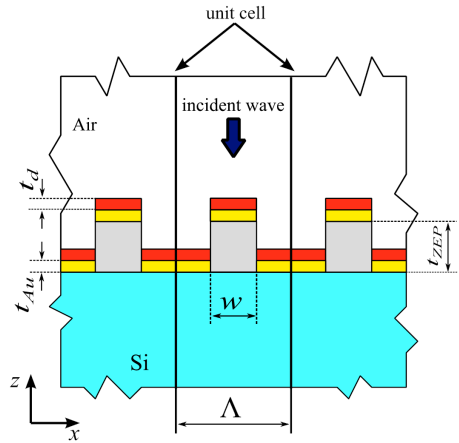

Fig. 1. Schematic of the hybrid metallo-dielectric 1-D nanograting (cross section view) considered in the experimental study, consisting of a gold film of thickness $t_{Au}=30nm$ (yellow layers) deposited on a 1-D patterned ZEP layer of thickness $t_{ZEP}=370nm$ and RI $n_{ZEP}=1.54$ with period $\Lambda=1500nm$ and $DC=w/\Lambda=0.25$ (light grey layers), backed by a silicon substrate ($n_s=3.4$). Dielectric (SiO$_2$, $n_d=1.45$) overlays (red layers) of thickness $t_d=30nm$ and $60nm$ are deposited on the top surface of the nanograting in order to estimate its surface sensitivity. Also shown is the unit cell considered in the numerical simulations, as well as the illumination from air.

Accordingly, in what follows, we numerically and experimentally study the *surface* sensitivity of RA-based metallic nanograting sensors, starting with nanosized overlays mimicking local RI changes. We show that, in this case, these devices exhibit a rather low (up to two orders of magnitude below the bulk figures) sensitivity, which is however in line with what observed in other grating-assisted sensing platforms [13]. Interestingly, we show that such residual sensitivity is not attributable to the RA wavelength shift, but rather to a strictly connected SPP effect. As a consequence, we found that the grating duty-cycle (DC) value that optimizes the surface sensitivity is generally different from that considered in the bulk case in order to maximize the RA visibility.

Moreover, we show how, for increasing overlay thickness (up to wavelength-sized values comparable with the typical sensing-volumes of plasmonic modes), this sensitivity approaches as an ultimate limit the RA bulk sensitivity, which represents an upper-bound for grating-assisted SPP sensitivities.

## 2. Experimental study on the RA surface sensitivity

Compatibly with our fabrication facilities [13], but without loss of generality, we begin our analysis by considering the 1-D metallo-dielectric structure schematically depicted in Fig. 1. It essentially consists of a dielectric (ZEP) 370 nm thick layer backed by a silicon substrate and patterned with a 1-D lattice with period $\Lambda = 1500 nm$ and duty-cycle $DC = w/\Lambda = 0.25$. A 30$nm$ thick film of gold is deposited on both the ridges and grooves of the dielectric grating.

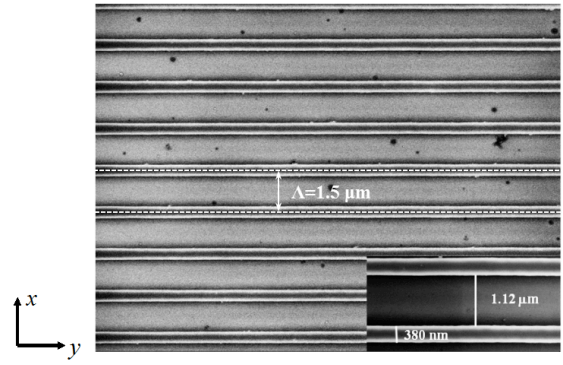

Fig. 2. SEM image (top view) of the fabricated device, with indication of its period $\Lambda = 1.5 \mu m$. The inset shows a magnified view and the widths of the ridges (380 nm) and grooves (1.12μm).

### 2.1. Fabrication

The nanograting was fabricated via electron beam lithography (EBL) followed by gold sputtering. The first step was the deposition of 370$nm$ of e-beam positive resist (ZEP 520A, Zeon Chemicals) via spin coating onto a cleaned square silicon substrate (10×10 mm$^2$). After spinning, the sample was baked on a hot plate for 3 minutes at 180°C. The periodic pattern was then defined on an area of 400×400 μm$^2$ by using a Raith 150 EBL system with a voltage of 20kV and a dose of 55 μC/cm$^2$. The grating lines were fabricated by using the Fixed Beam Moving Stage mode, which yields high precision and flexibility in defining thin but extended lines (or paths) with no field stitching boundaries. This operation allows the fabrication of well-defined line widths with no need to rely on a defocused beam. After the exposure, the sample was developed in a ZED N50 solution (Zeon Chemicals) for 55 sec. and rinsed in isopropanol. A nitrogen flow was subsequently utilized to remove the possible resist residue. The resulting dielectric grating was then coated with a 30 nm thick gold layer by DC magnetron sputtering in a vacuum system at a base pressure of $8\times10^{-5}$ Torr. The 30 and 60 nm thick SiO$_2$ layers were laid on top of the structure by RF sputtering at a pressure of 3mTorr with a rate of 3.3 Å/sec.

A scanning electron microscope (SEM) image (top view) of the fabricated device is shown in Fig. 2. The inset contains a magnified detail of the grating, showing the high degree of uniformity down to few nanometer length-scales.

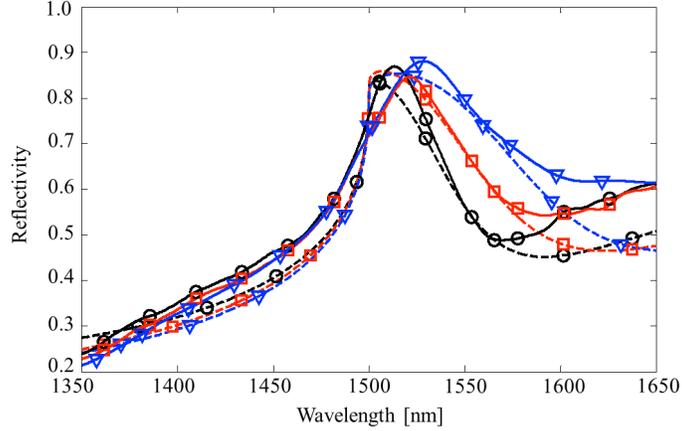

Fig. 3. Comparison among measured (solid curves with markers) and numerical (dashed curves with markers) reflectivity spectra pertaining to the device in Figs. 1 and 2, for different values of thickness $t_d$ of the SiO$_2$ overlay ($n_d$=1.45): $t_d = 0$ (i.e., no overlay; black curves with circles), $t_d = 30\,nm$ (red curves with squares), and $t_d = 60\,nm$ (blue curves with triangles).

## 2.2. Characterization

The experimental setup for the out-of-plane reflectivity measurements comprises a broad-band white light source directly coupled to a standard fiber-based reflection probe (AVANTES - Standard RP). Such probe illuminates the sample (placed on a XYZ positioning stage with 10 µm absolute on-axis accuracy) via an optical fiber bundle with six outside fibers in a ring-shaped configuration, and collects the reflected light by means of a central fiber (with a collection spot diameter ~350 µm) connected to an optical spectrum analyzer (with 0.4 nm resolution). The fibers' numerical aperture is 0.22. A 9 mm distance between the reflection probe and the sample was chosen as a suitable tradeoff between a reasonable signal-to-noise ratio and the fulfillment of the paraxial conditions (normal incidence). Finally, the measured reflectivity spectra were normalized with respect to the reference response of an unpatterned gold film (on the same substrate, but away from the grating area).

## 2.3. Results and discussion

Figure 3 shows the measured reflectivity spectrum (black-solid curve) of the fabricated device, for normal incidence (i.e., $\theta_i = 0$) from air. In accord with the theoretical predictions, the response exhibits a rather steep rising front around the 1st-order RA wavelength (pertaining to air) $\lambda_{R,air}^{(1)} = \Lambda = 1500\,nm$ [cf. Eq. (1)], at which the first ($m = \pm 1$) diffracted order reaches the grazing condition [5]. Also shown (black-dashed curve) in the same figure is the numerical response computed via a Rigorous Coupled Wave Analysis (RCWA) algorithm [14], assuming as RI for ZEP and silicon $n_{ZEP} = 1.54$ and $n_s = 3.4$, respectively, and relying on a Lorentz-Drude model [15] to account for the dispersion of gold. The grating period, DC and thickness were set to their nominal design values (1500 $nm$, 0.25 and 370 $nm$, respectively), and normal-incidence conditions were assumed. For a meaningful comparison, since no polarization control is implemented in our measurement setup (cf. Section 2.2), the numerical response was obtained by averaging the contributions of the transverse-electric (TE) and transverse-magnetic (TM) polarizations (i.e., electric field parallel and orthogonal to the grating lines, respectively). Moreover, consistently with the experiment, the response was also normalized with respect to the reflectivity response of a gold mirror. A good agreement between experimental and numerical results can be observed although, around the reflectivity peak, the experimental response exhibits a smoother profile by comparison with the numerical counterpart. This may be attributable to fabrication tolerances as well as slight deviations from the normal incidence condition in our experimental setup. It should also be noticed that our simulations assume an idealized conformal deposition of the $SiO_2$ overlay (i.e., with same thickness on both the grating ridges and grooves) which is likely different from that actually attained in the fabricated device. For this configuration, a bulk sensitivity

$$S_t = \frac{\lambda_{R,d}^{(1)} - \lambda_{R,air}^{(1)}}{n_d - 1} = \Lambda = 1500 \ (nm/RIU), \tag{2}$$

is theoretically predicted [5,9], with $\lambda_{R,d}^{(1)} = n_d \Lambda$ denoting the 1st-order RA wavelength pertaining to the dielectric [cf. Eq. (1)] of RI $n_d$.

In order to experimentally estimate the *surface* sensitivity of our metallic nanograting, we deposited on top of the structure (cf. Fig. 1) nanosized dielectric layers of $SiO_2$ ($n_d = 1.45$) of thickness $t_d = 30nm$ and $60nm$. The corresponding experimental and numerical responses are also shown in Fig. 3 as red ($t_d = 30nm$) and blue ($t_d = 60nm$) curves. The steep rising front around $\lambda_{R,air}^{(1)} = 1500nm$ practically does not undergo sensible changes in the presence of the thin $SiO_2$ overlays. The falling front, instead, broadens up, and a slight red-shift of the reflectivity peak (specifically, $8nm$ and $15.2nm$ for $t_d = 30nm$ and $60nm$, respectively) can be observed when the overlay thickness is increased. Also in this case, the experimental responses match the numerical ones, demonstrating a good agreement. In essence, the main effect of the overlay deposition is a broadening of the RA peak. This may be quantified by looking at the 80% linewidth of the reflectivity peak, which increases from 42 nm (no overlay), to $57nm$ ($t_d = 30nm$) and $73nm$ ($t_d = 60nm$), resulting in a maximum broadening up to 73.8%. In addition to the resonance linewidth broadening, also a red shift of the reflection peak can be observed. Consequently, we estimated the surface sensitivity by monitoring the wavelength shift of the reflectivity peak from $\lambda_{air}^{(peak)}$ to $\lambda_d^{(peak)}$ (i.e., in the absence and presence of the overlay, respectively), viz.,

$$S = \frac{\lambda_d^{(peak)} - \lambda_{air}^{(peak)}}{n_d - 1} \ (nm/RIU), \tag{3}$$

for the two overlay thickness values and $n_d = 1.45$.

Table 1. Numerical and experimental surface sensitivities [estimated via Eq. (3)] pertaining to the device in Figs. 1 and 2, for two values of thickness of the $SiO_2$ overlay. Also given, as a reference, is the theoretical estimate of the bulk (i.e., $t_d \to \infty$) sensitivity [cf. Eq. (2)].

| $t_d$ [nm] | Sensitivity $S$ [nm/RIU] | |
|---|---|---|
| | Numerical | Experimental |
| 30 | 8 | 17.77 |
| 60 | 25.33 | 33.77 |
| ∞ (bulk) | 1500 | - |

The experimental and numerical results are summarized in Table 1. In spite of the moderate discrepancies (attributable to the previously mentioned nonidealities and imperfections), the surface sensitivities are found to be about two orders of magnitude lower than the bulk sensitivity (i.e., 1500 *nm*/RIU) which depends only on the grating period. Nevertheless, the above figures are consistent with the values observed in other grating-assisted sensing schemes. For instance, in a previous study on hybrid metallo-dielectric photonic/plasmonic crystals [13], we experimentally verified wavelength shifts of ~ 4*nm* in the presence of 15*nm* of the same (SiO$_2$) dielectric overlay. Assuming a linear dependence on the overlay thickness, this corresponds to surface sensitivity values in line with our observations above.

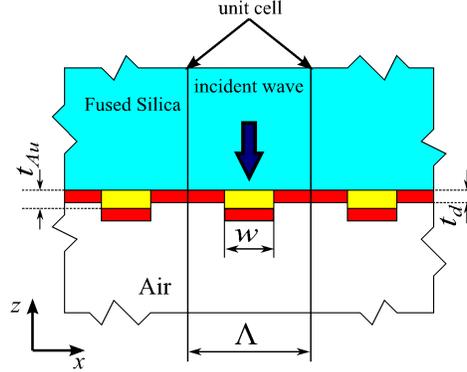

Fig. 4. Schematic of the sensor configuration considered in the numerical study, constituted by a linear gold (RI= $n_{Au}$, yellow layers) grating in the *x,z* plane, with period $\Lambda = 900nm$, thickness $t_{Au} = 30nm$, and $DC = w/\Lambda$, laid on a fused-silica ($n_s = 1.45$) substrate. A dielectric-analyte ($n_d = 1.33$, red layers) overlay of thickness $t_d$, surrounded by air (RI=1), is assumed. Also shown is the unit cell considered in the numerical simulations, as well as the illumination from the substrate.

## 2.4. Remarks on the general validity of our results

It should be stressed that the above results and related observations are not limited to the specific metallo-dielectric nanograting structure in Fig. 1 (which was chosen only for fabrication convenience), but are instead quite general, and hold for different nanograting configurations. To prove this point, in what follows, we proceed with a numerical study the RI surface sensitivity of the device considered in [9]. Besides confirming the general character of our findings, this will also allow us to consider, in the rest of the discussion, a structurally simpler configuration which greatly facilitates a clean-cut interpretation of the underlying phenomenologies. The analyzed device, schematically shown in Fig. 4, consists of a gold (RI= $n_{Au}$) linear grating characterized by same period $\Lambda = 900nm$, $DC = w/\Lambda$, and thickness $t_{Au} = 30nm$, laid on a fused-silica substrate (RI= $n_s = 1.45$). Differently from our initial configuration in Fig. 1, a fiber-optic-type illumination (i.e., impinging from the substrate) is considered here. Moreover, in view of the fused-silica substrate, a different dielectric overlay ($n_d = 1.33$) is considered instead of SiO$_2$.

Figure 5(a) shows the numerical reflectivity spectra for overlay thickness values $t_d = 0$, $t_d = 10nm$, and $t_d = 20nm$. As expected, the spectral responses exhibit trend similar to the previous case: the steep rising front at the 1st-order RA wavelength $\lambda_{R,air}^{(1)} = 900nm$ does not exhibit a sensible modification, while the falling front broadens up, thereby inducing a slight red-shift of the reflectivity peak.

## 3. Nature of the surface sensitivity

For a deeper understanding of the observed surface sensitivity, it is insightful to look at the numerical reflectivity spectra pertaining to the TM and TE polarizations, shown in Figs. 5(b) and 5(c), respectively.

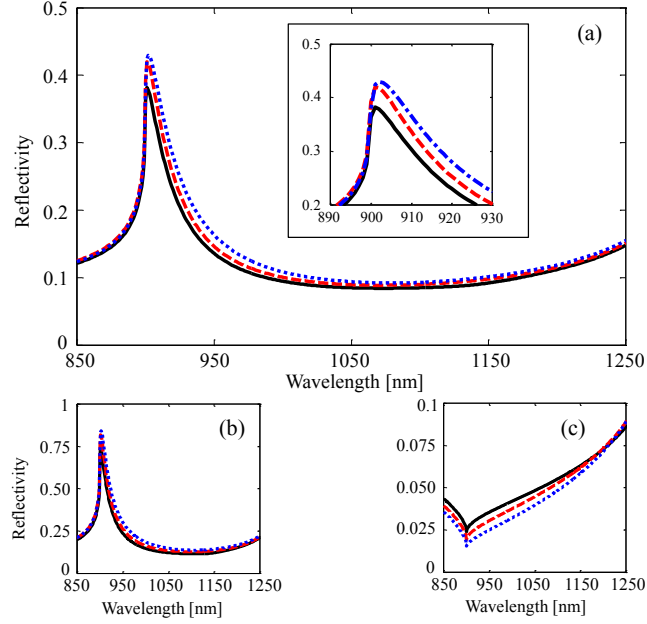

Fig. 5. (a) Numerical reflectivity spectra (magnified in the inset) pertaining to the device in Fig. 4, with $DC=0.35$, and for different values of thickness $t_d$ of the dielectric ($n_d=1.33$) overlay: $t_d=0$ (i.e., no overlay; black curves), $t_d=10nm$ (red curves), and $t_d=20nm$ (blue curves). The responses are obtained by averaging those pertaining to the TM and TE polarizations, shown separately in (b) and (c), respectively.

Such spectra, which were averaged in Fig. 5(a), turn out to be quite different. First, the dynamic range is quite different (note the different scale). Moreover, while both responses exhibit a quite visible spectral feature (peak and dip, respectively) around the 1st-order RA wavelength $\lambda_{R,air}^{(1)}=900nm$, the broadening front that is responsible for the actual (albeit small) surface sensitivity observed in Fig. 5(a) is visible only for the TM-polarization [cf. Fig. 5(b)]. Conversely, in the reflectivity spectra pertaining to the TE-polarization [cf. Fig. 5(c)], the presence of the overlay induces only a slight offset on the reflectivity scale, without any shift of the spectral dip. From this evidence, it is possible to recognize that the small residual surface sensitivity observed is not attributable to a genuine RA wavelength shift, but rather to the interplay with other phenomena that occur only for the TM-polarization. The most intuitive interpretation of such interplay would involve the excitation of a *plasmonic* (SPP) mode at a wavelength close to $\lambda_{R,air}^{(1)}$, which is responsible for the falling front of the spectral reflectivity peak [16], and cannot occur for the TE-polarization.

This even better illustrated in Fig. 6, which compares the reflectivity spectra and relevant field-maps in the presence and absence of a thicker ($t_d=150nm$) overlay, for the TM-polarization. Basically, in the absence of the overlay [$t_d=0$, cf. Fig. 6(a)], only the steep rising front of the reflectivity peak is actually attributable to an RA, whereas the falling front exhibits a plasmonic character. This is evident from the comparison between the *extended* field distributions at the RA wavelength [Fig. 6(b)] and the localized character observed at slightly higher wavelengths [Fig. 6(c)]. Our conclusions are further supported by the responses in the presence of the overlay. From the reflectivity spectrum shown in Fig. 6(d), we observe that the spectral feature at the 1st-order RA wavelength is clearly de-emphasized but still visible at $\lambda_{R,air}^{(1)}=900nm$ in the form of a cusp, with associated field distribution of *extended* character [Fig. 6(e)]. On the other hand, the main spectral peak (red-shifted at 1010 nm) is now associated with a *genuine* plasmonic mode exhibiting a strong localization at the metal-dielectric interface [Fig. 6(f)].

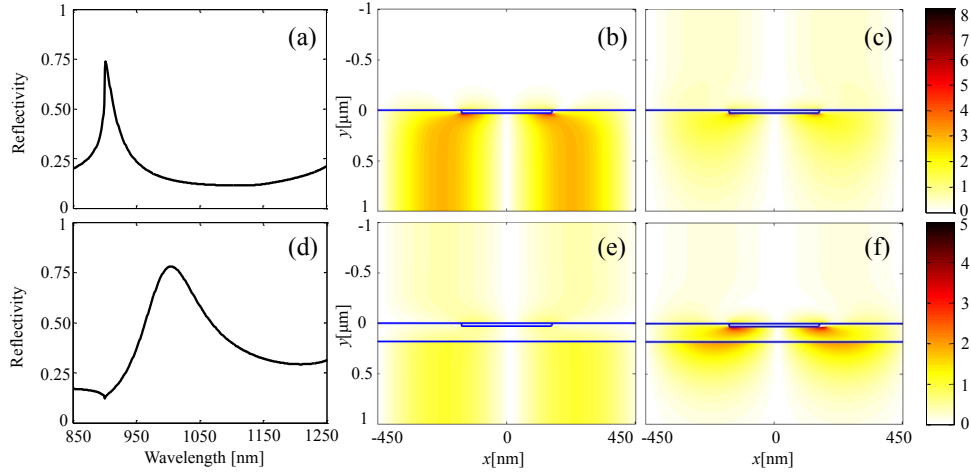

Fig. 6. (a), (d) Reflectivity spectra (TM polarization) pertaining to the device in Fig. 4, with $DC=0.35$, in the absence and presence of a dielectric ($n_d=1.33$) overlay of thickness $t_d=150nm$, respectively. (b), (c) Numerically-computed electric-field (z-component) magnitude maps within a unit cell, in the absence of the overlay, at wavelengths 900nm and 926 nm, respectively. (e), (f) Same of (b), (c), but in the presence of the overlay, at wavelengths 900 nm and 1010 nm, respectively. The thick horizontal line indicates the gold metallization at the interface between the substrate ($z<0$) and analyte ($z>0$) regions.

To sum up, our observations above indicate that RAs are *per se* completely insensitive to surface modifications of the ambient RI. The residual small sensitivity observed is actually attributable to their interplay with SPP modes excited at nearby wavelengths.

Within this framework, we recall that the particular choice of the grating DC in [9] (and also utilized here) was the result of an optimization process aimed at maximizing the visibility of the RA spectral feature in the form of a sharp peak, rather than as a moderate jump or cusp which would be difficult to detect experimentally (cf. Fig. 2 in [9]). Our interpretation above suggests that such optimization basically corresponds to exciting an SPP close to an RA wavelength. However, in view of the actual plasmonic nature of the residual surface sensitivity, and recalling that the grating *DC* also affects the SPP field-decay in the direction orthogonal to the metal-dielectric interface (which determines the sensing volume and hence the surface sensitivity), it turns out that the *DC* value that optimizes the RA visibility may not necessarily be the most convenient choice here. Figure 7 shows the surface sensitivity [computed via Eq. (3)] as a function of the overlay thickness $t_d$ varying within the range 40-200*nm*, for *DC*=0.35, 0.45 and 0.55. It is evident that the value *DC*=0.35, chosen in [9] to maximize the RA visibility (in their bulk configuration), does not guarantee the best surface sensitivity, which is instead attained for larger *DC* values. Further improvements of the surface sensitivity based on plasmonic effects are also possible via the deposition of functional high refractive index overlays at nanoscales (so as to enhance the field content at the metal-dielectric interface) [17], or by selecting sensing configurations involving localized-surface-plasmon-resonance (LSPR) effects characterized by much stronger field localization at the metal interface [18].

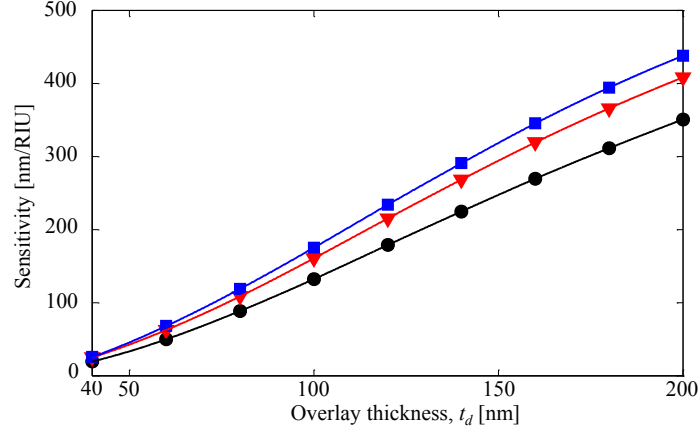

Fig. 7. Numerical surface sensitivity [estimated via Eq. (3)] pertaining to the device in Fig. 4, as a function of the thickness $t_d$ of the dielectric ($n_d = 1.33$) overlay, for different *DC* values: *DC*=0.35 (circles), *DC*=0.45 (triangles), and *DC*=0.55 (squares).

## 4. Ultimate sensitivity

It is now interesting to explore the connection between the RA and SPP-based sensing mechanisms that coexist in this type of nanograting devices. In particular, one might wonder whether both mechanisms essentially boil down to similar sensitivities when the associated sensing volumes are completely filled by the analyte.

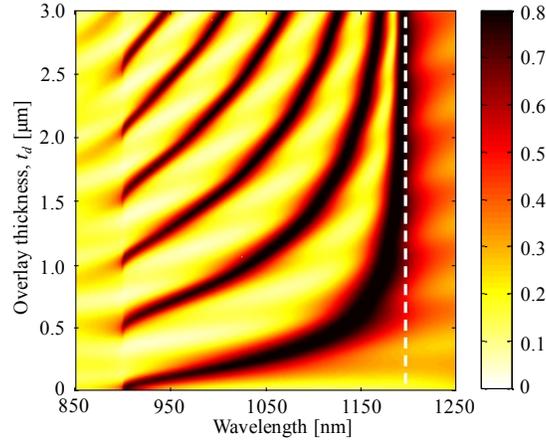

Fig. 8. Reflectivity spectra contour plot (TM polarization) of the structure in Fig. 5 for a dielectric ($n_d = 1.33$) with thickness $t_d$ varying within the range 0-3 *μm*. The white-dashed reference line corresponds to $\lambda_{R,d}^{(1)} = n_d \Lambda = 1197 nm$.

In order to gain some insight into this important aspect, we carried out an extensive parametric study by varying the dielectric overlay thickness. For the above nanograting, and TM-polarization, the contour plot in Fig. 8 shows, in false-color scale, the evolution of the reflectivity spectrum for increasing values of the dielectric ($n_d = 1.33$) overlay thickness $t_d$ (up to 3 *μm*), and $DC = 0.55$. We can observe a series of reflectivity peaks (dark ridges in the contour plot) that originate at $\lambda_{R,air}^{(1)} = \Lambda = 900 nm$ (corresponding to the 1st-order RA pertaining to air) and progressively red-shift, for increasing values of the overlay thickness, towards a limiting value $\lambda_{R,d}^{(1)} = n_d \Lambda = 1197 nm$ (corresponding to the 1st-order RA pertaining to the dielectric). These peaks are physically associated with bound modes supported by the structure, including the aforementioned SPP mode (which can be observed for arbitrarily thin overlays) and the growing number of guided-modes that the overlay can support as its thickness increases. Both SPP and guided (photonic) modes "break-down" at $\lambda_{R,d}^{(1)} = 1197 nm$, for which the wavevector associated to the diffractive wave needed for their excitation becomes purely imaginary [5].

The above results indicate that the transition between the two limit configurations (i.e., no analyte and bulk-analyte) is established, for increasing values of the overlay thickness, via the formation and progressive red-shift of spectral peaks associated with the discrete spectrum of bound (SPP and guided) modes that the structure can support, whose merging (for $t_d \to \infty$) eventually gives rise to the sharp spectral peak at the RA wavelength pertaining to the bulk dielectric.

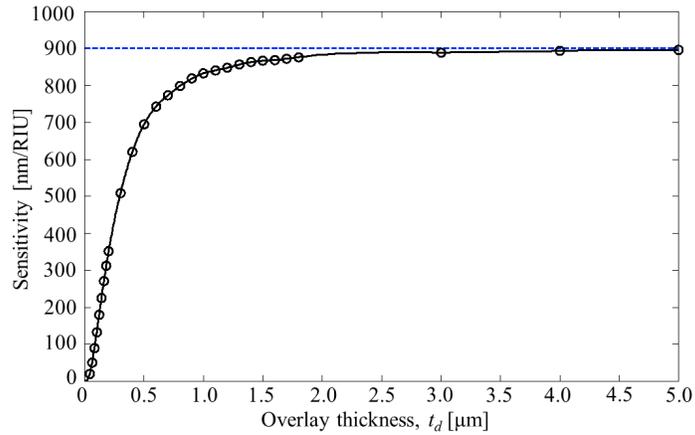

Fig. 9. Numerical surface sensitivity [estimated via Eq. (3)] pertaining to the device in Fig. 4, with $DC = 0.35$, as a function of the thickness $t_d$ of the dielectric ($n_d = 1.33$) overlay. The dashed horizontal line corresponds to the theoretical bulk-sensitivity limit $S_t = \Lambda = 900\ nm/RIU$.

Of particular interest for assessing the surface sensitivity is analysis of the SPP mode behavior. Figure 9 shows the sensitivity associated with the corresponding peak wavelength-shift [computed via Eq. (3)], as a function of the overlay thickness $t_d$. As it can be observed, the sensitivity initially exhibits a rather steep increase (of nearly two orders of magnitude from the very low levels previously observed in the presence of nanosized overlays) for thickness values up to 300 nm, and then approaches (for thickness values ~ 2µm) the theoretical RA bulk-sensitivity prediction $S_t = \Lambda = 900 nm/RIU$ [cf. Eq. (2)]. From the physical viewpoint, this ultimate sensitivity is consistent with the complete filling (by the analyte-overlay) of the wavelength-sized sensing volume of SPP modes.

## 5. Conclusions

In this paper, we have studied, numerically and experimentally, the surface sensitivity of RAs in metallic nanogratings. Our main results can be summarized as follows:

i. Although extremely sensitive when bulk analytes are considered, the spectral features associated to RAs in metal nanogratings are completely insensitive when local RI changes are considered. This finding severely curtails the application of RA-based technological platforms for chemical and label-free biosensing.

ii. The residual surface sensitivity observed (up to two orders of magnitude lower than the RAs bulk value) is essentially attributable to interplaying plasmonic (SPP) phenomena spectrally overlapping with RAs, and is in line with the figures observed in similar scenarios for other types of grating-assisted sensing schemes [13].

iii. Grating DC values that optimize the RA visibility in the presence of a bulk analyte are generally not optimal in connection with the (SPP-based) surface sensitivity.

iv. Finally, the ultimate surface sensitivity approaches the theoretical bulk-sensitivity (corresponding to the grating period) associated with RAs for wavelength-sized overlays (i.e., entirely filling the typical sensing volume of SPP modes).

The above results imply that when metal nanogratings are employed, the RA- and SPP-based sensing mechanisms essentially boil down to similar sensitivities, as long as the associated sensing volume is completely filled. In the intermediate regime (cf. Fig. 7), of interest for practical applications to chemical and biological sensing, particular care should be exercised in the choice of the grating materials and geometry (as well as the wavelength-range), so as to ensure a suitable plasmonic response. Possible improvements, currently under investigation, may be attained via the deposition of nanosized high-refractive-index overlays, or via the excitation of LSPRs.

Besides shedding light in the physical phenomena (and their interplay) underlying RA-based nanodevices, our results above also provide useful quantitative assessments for their applicability to label-free chemical and biological sensing.

**Acknowledgment**

The kind assistance of Mr. Salvatore Cifarelli (University of Sannio) for the spectral measurements is gratefully acknowledged.